\documentclass[aps, prd,reprint,superscriptaddress, keywords,pacs,nofootinbib,amsmath,amssymb,floatfix]{revtex4-2}

\usepackage{graphicx}
\usepackage{dcolumn}
\usepackage{bm}
\usepackage{hyperref}
\usepackage{xcolor}
\usepackage{tikz}
\usepackage{tikz-feynman}
\tikzfeynmanset{compat=1.1.0}

\begin{document}

\title{One-loop Renormalization of the Improved Energy–momentum Tensor in  Lattice QCD}

\author{Mushtaq Loan}
\email{Corresponding author}
\affiliation{Department of Physics and Mathematics, Sichuan University Pittsburgh  Institute (SCUPI), Chengdu, 61000, China}

\author{Nasser Demir}
\affiliation{Department of Physics, College of Science, Kuwait University, Safat, 1320, Kuwait}

\date{\today}

\begin{abstract}
We present the one-loop renormalization of the improved gluonic energy–momentum tensor (EMT) by employing a three-loop-improved clover discretization of the field-strength tensor 
in pure SU(3) lattice gauge theory. The renormalization factor is extracted  by matching the amputated two-gluon matrix element of the lattice energy–momentum tensor to the continuum 
$\overline{\mathrm{MS}}$ scheme. The one-loop contribution is separated into sail, operator-tadpole, and external-leg corrections, each expressed in terms of a minimal set of scalar Brillouin-zone integrals to obtain 
the explicit expression for the finite lattice coefficient  $\mathfrak{B}_{\mathrm{lat}}(u_{0})$ and the multiplicative renormalization factor $Z_{T}(u_{0})$ associated with the traceless spin-2 component of 
the energy–momentum tensor. A key result, obtained in this framework, is the clear distinction between two sectors: the spin-2 sector, governed by $Z_{T}$, and the scalar trace sector, which encompasses 
the Yang–Mills trace anomaly. The trace is determined by the scalar operator $F_{\rho\sigma}F_{\rho\sigma}$  and the Yang–Mills $\beta$ function, rather than by the spin-2  renormalization factor. 
The renormalized EMT modifies the normalization and short-distance behaviour of energy-density correlators through both traceless and scalar-channel contributions. 
Comparison with existing lattice thermodynamic data demonstrates that the improved operator accurately reproduces the expected temperature dependence of the trace anomaly 
and offers a systematically improvable framework for investigating the equation of state, gluon condensate, and transport coefficients in lattice QCD.

\end{abstract}

\maketitle

\section{Introduction}
\label{sec:introduction}
The energy--momentum tensor (EMT) is a central operator in quantum field theory. As the conserved current associated with spacetime translational invariance \cite{Freese2020,Freese2026}, 
its matrix elements encode the hadron’s momentum and spin fractions \cite{Yang2018,Alexandrou2020,Haghani2024}, its expectation value in thermal states defines the equation of state, and 
all thermodynamic metrics of  strongly-coupled matter \cite{Asakawa2014}, and encodes physically important quantities such as the energy density, momentum flow, thermodynamic 
observables, and gravitational form factors \cite{Cao2024,Cao2026}. In Yang-Mills theory and QCD, a correctly normalized EMT is also required for the nonperturbative study of the equation of 
state, transport coefficients, hadron and glueball matrix elements, and the trace anomaly.

On the lattice, however, the construction of the energy-momentum tensor (EMT) is nontrivial because continuous translational symmetry is explicitly broken to a discrete subgroup. 
The naïve discretization of the EMT operator leads to severe ultraviolet divergences—extreme fluctuations at very short distances—that require multiplicative and additive 
renormalization \cite{LuscherWeisz1985}. Unlike simpler fermion bilinears, the EMT mixes with other operators of equal or lower dimension under renormalization \cite{Caracciolo1990,Capitani1995}, 
and its renormalization constants depend on the lattice action and the specific discretization of the gauge and fermion fields. Moreover, the conservation of the EMT, 
which is exact in the continuum, is broken by lattice artifacts - that is, artifacts arising specifically from the discretization - making a direct extraction of physical 
matrix elements from lattice correlators a delicate task.

Two distinct frameworks have been developed to address these challenges. The first framework utilizes shifted boundary conditions to introduce a non-perturbative renormalisation 
scheme, thereby eliminating the need to explicitly construct composite operators on the lattice. Imposing twisted boundary conditions in the Euclidean time direction relates the partition function of the 
EM tensor to a derivative with respect to the shift parameter. Such a relationship facilitates the extraction of renormalized EMT matrix elements via a Ward identity \cite{Giusti2011,Giusti2015}. 
This approach has been successfully applied to the renormalisation of the EM tensor in pure Yang-Mills theory \cite{Giusti2015} and later extended to full QCD \cite{Brida2020}, to accurately reproduce the trace anomaly.
The second approach, the so-called gradient flow, uses a continuous smoothing transformation on gauge and fermion fields along a positive ‘flow time’ $t$. This technique reduces ultraviolet quantum 
fluctuations while preserving the essential long-distance physics of strong interactions. Suzuki first showed that a finite, renormalised energy-momentum tensor (EMT)—which measures the density 
and flow of energy and momentum in a field—could be built from flowed fields in pure gauge theory \cite{Suzuki2013}. This idea was later extended to include quarks \cite{Makino2014}, offering 
a practical, gauge-invariant EMT definition on the lattice (a discrete spacetime grid used in numerical simulations). This method has now become standard for lattice QCD thermodynamics, supporting 
high-precision calculations of the trace anomaly and the equation of state \cite{Asakawa2014}. These advances set a benchmark for perturbative calculations: one-loop lattice perturbation theory 
renormalisation factors—correction factors accounting for quantum effects at one order—should match the correct weak-coupling behaviour and can be directly compared with 
nonperturbative lattice data \cite{LepageMackenzie1993}.

Perturbative lattice perturbation theory complements these non-perturbative methods by providing crucial input on operator mixing coefficients and renormalization constants. 
Specifically, one-loop computations of the renormalization of the EMT have been carried out for a variety of lattice actions, \cite{Caracciolo1990,Capitani1995,CaraccioloMenottiPelissetto1992,Capitani2003}, 
serving as a necessary component in the development of improved actions as well as a consistency check. More recently, high-precision non-perturbative matching between the lattice EMT 
and continuum schemes like $\overline{\mathrm{MS}}$ \cite{Costa2021} has been made possible by the development of gauge-invariant renormalization schemes building on these 
foundations and allowing direct comparison with phenomenological extractions. 

The present work is situated at the interface between improved lattice perturbation theory and nonperturbative EMT renormalization. Our goal is to compute the one-loop renormalization of a 
tadpole-improved gluonic EMT built from a three-loop improved clover field-strength tensor, together with the tree-level Symanzik gauge action containing plaquette and rectangle terms. 
In this framework, the  one-loop matching factor is obtained by evaluating the two-gluon matrix element of the operator in lattice perturbation theory and converting 
it to a continuum renormalization scheme. The calculation naturally separates into sail, tadpole, and external-leg contributions, each of which contributes to the finite lattice coefficient 
entering the renormalization factor. Normalization ensures that the clover EMT has tree-level coefficients of order unity, making perturbative 
matching (e.g., to $\bar{MS}$) more stable and convergent. To elaborate, a central issue in this analysis is the distinction between the traceless spin-2 channel and the scalar 
trace channel. In particular, the multiplicative renormalization constant $Z_T$ extracted from the spin-2 projector fixes the normalisation of the traceless part of the EMT, while 
the trace anomaly is encoded in the scalar operator proportional to $F_{\rho\sigma}^{a}F_{\rho\sigma}^{a}$. Consequently, a complete renormalisation program must track 
both sectors: the perturbative matching of the spin-2 operator and the scalar-channel normalisation required to reproduce the EMT in the continuum limit.

The paper is organized as follows. We first define the tadpole-improved three-loop clover EMT 
to derive the relevant propagators and operator vertices in Landau gauge, including the two-, three-, and four-gluon EMT kernels. Next, we evaluate the 
one-loop sail, tadpole, and external-leg diagrams and reduce the projected amplitudes to a minimal basis of Brillouin-zone integrals. 
This yields the lattice finite coefficient \(\mathfrak B_{\rm lat}(u_0)\) and, through matching to the continuum scheme, the renormalization constant \(Z_T\). 
Finally, we discuss the trace anomaly, the scalar channel, and the comparison between the perturbative prediction and results obtained using 
shifted boundary conditions and gradient-flow methodologies.

\section{Theoretical Framework}
\subsection{Tadpole-improved Energy-Momentum Tensor on Lattice}
The action is discretized through the tadpole-improved  Symanzik gauge action  \cite{Lepage1993}
\begin{eqnarray}
S_{g}^{Imp} &= & \frac{\beta}{N_{C}}\left[\frac{c_{0}}{u_{0}^{4}} \sum_{plaq}\frac{1}{3} \mbox{Re Tr} (1-U_{plaq}) \right. \nonumber\\
& & \left. + \frac{c_{1}}{u_{0}^{6}} \sum_{rect.}\frac{1}{3} \mbox{Re Tr} (1-U_{rect})\right]
\end{eqnarray}
where the trace is over the colour indices and $\beta = 2N_{C}/g_{0}^{2}$  with $g_{0}$ being the bare coupling constant. For the Symanzik action, the coefficients are $c{_0}=5/3$ and $c_{1}=-1/12$.

The plaquette is defined as a function of the gauge links, and it given by
\begin{displaymath}
U_{\mu\nu} (x)= U_{\mu}(x)U_{\nu}(x+ a \hat{\mu})U^{\dagger}_{\mu}(x+a \hat{\nu}) U^{\dagger}_{\mu}(x)
\end{displaymath} 
where $\mu , \mu =0, \cdots 3$, $\hat{\mu}$ is the unit vector along the direction $\mu$, and $x$ is the space-time coordinate. 

\begin{figure}[!h]
\scalebox{0.55}{\includegraphics{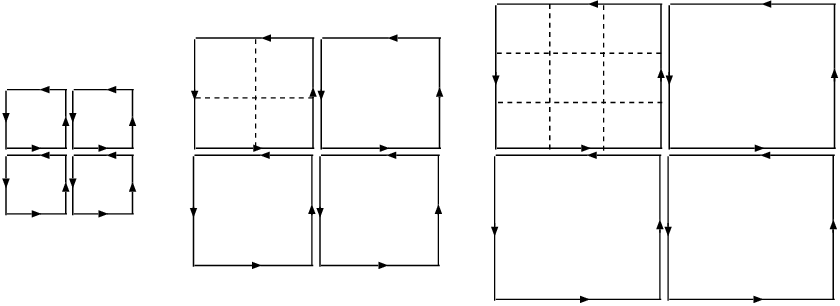}} 
\caption{\label{figa} The $1\times 1$, $2\times 2$, and $3\times 3$ loops are used to construct the clover term for the improved energy-momentum tensor.}
\end{figure}

The energy-momentum tensor of the gauge field theory has the form:
\begin{equation}
T_{\mu\nu}(x) = \frac{1}{g_{0}^{2}}\left[F_{\mu\alpha}^{a} F_{\mu\alpha}^{a} - \frac{1}{4}\delta_{\mu\nu}F_{\alpha\beta}^{a}F_{\alpha\beta}^{a}\right],
\label{eqn2}
\end{equation}
where the gluon field strength tensor is defined as
\begin{equation}
F_{\mu\nu}^{a}(x) = -\frac{i}{4a^{2}}\left[\bigg(Q_{\mu\nu}(x) - Q_{\mu\nu}^{\dagger}(x)\bigg)T^{a}\right],
\label{eqn2b}
\end{equation}
and
\begin{equation}
Q_{\mu\nu} = \frac{1}{4}\bigg(U_{\mu\nu}(x)+U_{-\nu\mu}(x)+U_{\nu-\mu}(x)+U_{-\mu-\nu}(x)\bigg)
\label{eqn3}
\end{equation}
is the sum of the four plaquette terms. The gluon field strength tensor has $O(a^{2})$ discretisation errors. To improve the discretisation of the gluon field strength tensor, 
we incorporate additional higher "clover" loops (Fig. \ref{figa}) in $F_{\mu\nu}$. In general, we define the following  improved field strength tensor 
\begin{eqnarray}
F_{\mu\nu}^{imp}(x) &= & k_{1}F_{\mu\nu}^{1\times 1}+ k_{2}F_{\mu\nu}^{2\times 2}+\frac{k_{3}}{2}\bigg(F_{\mu\nu}^{2\times 1}+
F_{\mu\nu}^{1\times 2}\bigg)\nonumber\\
& & \frac{ k_{4}}{2}\bigg(F_{\mu\nu}^{3\times 1}+ F_{\mu\nu}^{1\times 3}\bigg) + k_{5}F_{\mu\nu}^{3\times 3},
\label{eqn4}
\end{eqnarray}
where
\begin{eqnarray}
F_{\mu\nu}^{m\times n}(x) = - \frac{i}{4a^{2}}\left[\bigg(Q_{\mu\nu}^{m\times n}(x) - Q_{\mu\nu}^{\dagger m\times n}(x)\bigg)T^{a}\right]
\label{eqn5}
\end{eqnarray}
$k_{i}$ are the constant coefficients and $Q_{\mu\nu}^{m\times n}(x)$ corresponds to the sum of the four $m\times n$ loops in the clover formation. 

For computational efficiency, we consider a 3-loop improved field strength tensor ($k_{4}=k_{5}=0$) with mean field improved coefficients,
\begin{equation}
F_{\mu\nu}^{3L}(x) = \frac{3}{2u_{0}^{4}}F_{\mu\nu}^{1\times 1}(x)
-\frac{3}{20u_{0}^{8}}F_{\mu\nu}^{2\times 2}(x)+\frac{1}{90u_{0}^{12}}F_{\mu\nu}^{3\times 3}(x)
\label{eqn6}
\end{equation}
The improved energy-momentum tensor is then represented by
\begin{equation}
T_{\mu\nu}^{Imp}(x) = \frac{1}{g_{0}^{2}}\left[F_{\mu\alpha}^{3L} F_{\mu\alpha}^{3L}- \frac{1}{4}\delta_{\mu\nu}F_{\alpha\beta}^{3L}F_{\alpha\beta}^{3L}\right].
\label{eqn6b}
\end{equation}

\section{One-loop Renormalization of $T_{\mu\nu}^{Imp}(x)$}
\label{sec:1A}
The improved EMT on the lattice, $T_{\mu\nu}^{Imp.}$, is a composite operator. 
When inserted into correlation functions, it will have  UV divergences from point-splitting regularization, operator mixing with lower-dimensional operators, and lattice artifacts from broken continuum symmetries. 
The renormalized operator $T_{\mu\nu}^{R, Imp.}$ is related to its bare counterpart  by
 \begin{equation}
T_{\mu\nu}^{R}(x) = Z_{T} (g^{2}_{0})\bigg(T^{Imp}_{\mu\nu}(x) -\langle T^{Imp}_{\mu\nu}(x)\rangle \bigg).
\end{equation}
The subtraction of the vacuum expectation value, $\langle T^{Imp}_{\mu\nu}(x)\rangle$ is essential as it non-perturbatively removes the power divergent mixing $\approx \delta_{\mu\nu}/a^{2}$. 

\subsection{Propagator and $n$-gluon EMT Vertices}

The tree-level gluon propagator is obtained from the quadratic terms in the action. For each loop (plaquette or rectangle) the second order term is proportional to the lattice discretization of 
$(\partial_{\mu}A_{\nu}-\partial_{\nu}A_{\mu})^{2}$ plus higher-derivative discretization corrections.  
The contribution from the total quadratic action, after  expanding and Fourier transforming  in momentum space has the  form
\begin{displaymath}
S^{(2)} = \frac{1}{2}\sum_{k} A_{\mu}^{a}(-k)G_{\mu\nu}^{ab}(k)A_{\nu}^{b},
\end{displaymath}
where 
\begin{displaymath}
\hat{k}_{\mu} = \frac{2}{a} \sin \bigg(\frac{ak_{\mu}}{2}\bigg), \hspace{1.0cm}  \hat{k}^{2} = \sum_{\mu}\hat{k}^{2}_{\mu}.
\end{displaymath}
The quadratic kernal has the usual transverse and logitudinal decomposition
\begin{displaymath}
G_{\mu}(k) = A(k) P_{\mu}^{T}(k)+B(k)P_{\mu\nu}^{L}(k),
\end{displaymath}
where the functions $A(k)$ and $B(k)$ are simply the transverse and longitudinal eigenvalues of the quadratic (inverse–propagator) kernel $G_{\mu\nu}(k)$:
\begin{displaymath}
P_{\mu\nu}^{T}(k) = \delta_{\mu\nu}- \frac{\hat{k}_{\mu}\hat{k}_{\nu}}{\hat{k}^{2}}, \hspace{1.0cm} P_{\mu\nu}^{L}(k) = \frac{\hat{k}_{\mu}\hat{k}_{\nu}}{\hat{k}^{2}}.
\end{displaymath}
The functions $A(k)$ and $B(k)$  are obtained by projecting $G_{\mu\nu}$  onto the transverse/longitudinal subspaces:
\begin{eqnarray}
A(k) & = & \frac{1}{d-1}\mbox{Tr}\left[G(k)P^{T}(k)\right]=\frac{1}{d-1}P_{\mu\nu}^{T}(k)G_{\mu\nu}(k),\nonumber\\
B(k) & = &  \mbox{Tr}\left[G(k)P^{L}(k)\right]= P_{\mu\nu}^{L}(k)G_{\mu\nu}(k)= \frac{\hat{k}_{\mu}G_{\mu\nu}(k)\hat{k}_{\nu}}{\hat{k}^{2}},\nonumber
\end{eqnarray}
for $d=4$  Eucilidean lattice.\\

On the lattice, gauge invariance implies that the quadratic gauge kernel \(G_{\mu\nu}(k)\) has a null longitudinal mode,
\begin{displaymath}
\tilde{k}_\mu G_{\mu\nu}(k)=0,
\end{displaymath}
and is therefore not directly invertible. To define the gluon propagator, one adds the lattice gauge-fixing term
\begin{displaymath}
S_{\mathrm{gf}} = \frac{1}{2\xi}\sum_x\big(\nabla^-_\mu A_\mu^a(x)\big)^2,
\end{displaymath}
which contributes in momentum space
\begin{displaymath}
G_{\mu\nu}^{(\mathrm{gf})}(k)=\frac{1}{\xi}\,\hat{k}_\mu\hat{k}_\nu.
\end{displaymath}
The full quadratic kernel is then
\begin{displaymath}
G_{\mu\nu}^{(\mathrm{tot})}(k) = G_{\mu\nu}^{(0)}(k)+\frac{1}{\xi}\hat{k}_\mu\hat{k}_\nu,
\end{displaymath}
with \(G_{\mu\nu}^{(0)}(k)\) the gauge-invariant part coming from the Symanzik action. For an exactly gauge-invariant quadratic kernel, the 
longitudinal eigenvalue vanishes, while the transverse eigenvalue takes the form
\begin{displaymath}
A(k)=\hat{k}^2 K_T(k;u_0),
\end{displaymath}
where \(K_T(k;u_0)\) is the improvement factor induced by the plaquette and rectangle terms.

To enforce transversality of the propagator and remove the longitudinal modes, we choose the Landau gauge $(\xi \rightarrow 0)$. 
The tree-level gluon propagator in Landau gauge is therefore the inverse of the transverse kernel:
\begin{equation}
D_{\mu\nu}^{ab}(k;u_{0})|_{\mbox{Landau}}  =  \delta^{ab}\bigg(\delta_{\mu\nu}- 
\frac{\hat{k}_{\mu}\hat{k}_{\nu}}{\hat{k}^{2}}\bigg)\frac{1}{\hat{k}^{2} K_{T}(k;u_{0})},
\label{eqp}
\end{equation}
where the continuum-like transverse operator $(\hat{k}^{2}\delta_{\mu\nu}- \hat{k}_{\mu}\hat{k}_{\nu})$  receives the leading 
scalar transverse eigenvalue factor
\begin{equation}
K_{T}(k;u_{0}) =\left[ \tilde{c}_{0}+\tilde{c}_{1}a^{2}\hat{k}^{2}+O(a^{4}\hat{k}^{2}\right],
\label{eqnKT}
\end{equation}
with  the effective coefficients in the transverse kernel given by:
\begin{displaymath}
\tilde{c}_{0} = \bigg(\frac{c_{0}}{u_{0}^{4}}+\frac{8c_{1}}{u_{0}^{6}}\bigg), \hspace{1.0cm} \tilde{c}_{1} = -\frac{c_{1}}{u_{0}^{6}}.
\end{displaymath}

To calculate the EMT 2-,3-, and 4-gluon vertices, we expand  improved field strength, $F_{\mu\nu}^{n\times n}$, for each clover loop size $n=1,2,3$
\begin{displaymath}
F_{\mu\nu}^{n\times n} = F_{\mu\nu}^{(1), n\times n} +g_{0}F_{\mu\nu}^{(2), n\times n}+g_{0}^{2}F_{\mu\nu}^{(3), n\times n}+O(g_{0}^{3}).
\end{displaymath}
Therefore,
\begin{displaymath}
F_{\mu\nu}^{3L} = F_{\mu\nu}^{(1), 3L} +g_{0}F_{\mu\nu}^{(2), 3L}+g_{0}^{2}F_{\mu\nu}^{(3), 3L}+ O(g_{0}^{3}),
\end{displaymath}
with
\begin{displaymath}
F_{\mu\nu}^{(r), 3L} =\frac{3}{2u_{0}^{4}} F_{\mu\nu}^{(1), 1\times 1} - \frac{3}{20u_{0}^{8}}F_{\mu\nu}^{(2), 2\times 2}+\frac{1}{90u_{0}^{12}}F_{\mu\nu}^{(3), 3\times 3}.
\end{displaymath}
The EMT operator then expands as
\begin{equation}
T_{\mu\nu}^{Imp} = T_{\mu\nu}^{(2)}+g_{0}T_{\mu\nu}^{(3)}+g_{0}^{2}T_{\mu\nu}^{(4)}+O(g_{0}^{3}),
\label{eqT}
\end{equation}
where
\begin{eqnarray}
T_{\mu\nu}^{(2)}  & = & \frac{1}{g_{0}^{2}}\left[F_{\mu\alpha}^{(1),3L}F_{\nu\alpha}^{(1),3L} -\frac{1}{4}\delta_{\mu\nu}F_{\alpha\beta}^{(1),3L}F_{\alpha\beta}^{(1),3L}\right],\nonumber\\
T_{\mu\nu}^{(3)}  & = & \frac{2}{g_{0}}\left[F_{\mu\alpha}^{(1),3L}F_{\nu\alpha}^{(2),3L} -\frac{1}{4}\delta_{\mu\nu}F_{\alpha\beta}^{(1),3L}F_{\alpha\beta}^{(2),3L}\right],\nonumber\\
T_{\mu\nu}^{(4)}  & = & F_{\mu\alpha}^{(2),3L}F_{\nu\alpha}^{(1),3L} +2F_{\mu\alpha}^{(1),3L}F_{\nu\alpha}^{(3),3L}\nonumber\\
& & 
-\frac{1}{4}\delta_{\mu\nu}\bigg(F_{\alpha\beta}^{(2),3L}F_{\alpha\beta}^{(2),3L}+F_{\alpha\beta}^{(1),3L}F_{\alpha\beta}^{(3),3L}\bigg),
\label{eqVer}
\end{eqnarray}
generate, respectively, the 2-gluon, 3-gluon, and 4-gluon EMT vertices. In the momentum space, with the  standard lattice momenta
\begin{displaymath}
\hat{p}_{\mu} = \frac{2}{a}\sin (ap_{\mu}/2), \hspace{0.15cm} \tilde{p}_{\mu} = \sin (ap_{\mu}),  \hspace{0.15cm}  c_{\mu}(p) = \cos (ap_{\mu}/2),
\end{displaymath} 
 the linearised clover field strength for $n\times n$ loop  can be written as 
\begin{displaymath}
F_{\mu\nu}^{(1),n,a} = iR_{n}(p_{\mu})R_{n}(p_{\nu})(\hat{p}_{\mu}A_{\nu}-\hat{p}_{\nu}A_{\mu}),
\end{displaymath}
where the path sum factor $R_{n}(p_{\mu})$ is 
\begin{displaymath}
R_{n}(p_{\mu}) = \frac{\sin (nap_{\mu}/2)}{\sin (ap_{\mu}/2)}, 
\end{displaymath}
with $p_{\mu}$ as the external four-momentum carried by the EMT insertion matrix element. The plaquette-size kernal is defined by
\begin{displaymath}
\Omega_{n}^{\mu\nu}(p) =R_{n}(p_{\mu})R_{n}(p_{\nu})
\end{displaymath}
and the 3-loop ($3L$) improved kernal is then given by
\begin{displaymath}
F_{\mu\nu}^{(1), 3L}(p)=i\Omega_{3L}^{\mu\nu}(p;u_{0})(\hat{p}_{\mu}A_{\nu}-\hat{p}_{\nu}A_{\mu}),
\end{displaymath}
with
\begin{eqnarray}
\Omega_{3L}^{\mu\nu}(p;u_{0})& =& \sum_{n=1}^{3}c_{n}^{(F)}\Omega_{n}^{\mu\nu}(p)\nonumber\\
& =& \frac{3}{2u_{0}^{4}} \Omega_{1}^{\mu\nu} - \frac{3}{20u_{0}^{8}} \Omega_{2}^{\mu\nu}+\frac{1}{90u_{0}^{12}} \Omega_{3}^{\mu\nu}.\nonumber
\end{eqnarray}
Writing $T^{(2)}=\frac{1}{2}A\Lambda_{T}^{(2)}A$, the amputated 2-gluon operator vertex is then given by
\begin{equation}
\Lambda_{\mu\nu;\rho\sigma}^{ab, (2)}(p;u_{0}) = \delta^{ab} \frac{1}{g_{0}^{2}}\Omega_{3L}^{\rho\alpha}(p;u_{0})\Omega_{3L}^{\sigma\beta}(p;u_{0})Q_{\mu\nu;\rho\sigma}^{\alpha\beta}(p),
\end{equation}
where $Q$ is the standard gluonic EMT structure with lattice momenta $p_{\mu}, p_{\nu}$ and Lorentz indices $\rho, \sigma$
\begin{eqnarray}
Q_{\mu\nu;\rho\sigma}^{\alpha\beta}(p) & = & (\delta_{\mu\rho}\hat{p}_{\alpha}-\delta_{\mu\alpha}\hat{p}_{\rho}) (\delta_{\nu\rho}\hat{p}_{\alpha}-\delta_{\nu\alpha}\hat{p}_{\rho}) +(\rho \leftrightarrow \sigma)\nonumber\\
& & - \frac{1}{4} \delta_{\mu\nu}\left[  (\delta_{\gamma\rho}\hat{p}_{\alpha}-\delta_{\gamma\alpha}\hat{p}_{\rho}) (\delta_{\gamma\sigma}\hat{p}_{\beta}-\delta_{\gamma\beta}\hat{p}_{\sigma}) \right. \nonumber\\
& & \left.+(\rho \leftrightarrow \sigma)\right].
\end{eqnarray}
To obtain the EMT 3-gluon vertex,  we need to evaluate the quadratic part of the clover field strength, $F^{(2),n}$ in the second equation of  (\ref{eqVer}).  
For any Wilson loop, the quadratic term in the anti-Hermitian part can be written in momentum space as
\begin{displaymath}
F_{\mu\nu}^{(2),a,n} = \frac{1}{2}f^{abc}\int_{BZ}\frac{d^{4}q}{(2\pi)^4}\mathcal{S}_{\mu\nu;\rho\sigma}^{a}(p,q)A_{\rho}^{b}(q)A_{\sigma}^{c}(p-q),
\end{displaymath}
where, for the average $n\times n$ clover loop, the kernal $\mathcal{S}$ is obtained by summing over pair of link insertions on the loop and, after antisymmetrizing, is given  by
\begin{eqnarray}
\mathcal{S}_{\mu\nu;\rho\sigma}^{(n)}(p,q) & = & \left[\delta_{\rho\mu}\delta_{\sigma\nu}C_{\mu\nu}^{(n)}(q,p-q) \right. \nonumber\\
 & & \left. - \delta_{\rho\nu}\delta_{\sigma\mu}C_{\mu\nu}^{(n)}(q,p-q)\right] +\mathcal{T}_{\mu\nu;\rho\sigma}^{(n)}.
\label{eqK}
\end{eqnarray}
The first two terms in Eq. {\ref{eqK}) are the continuum-like commutor core, and $\mathcal{T}$ contains the lattice transport correlations. The core factor $C_{\mu}$ and the transport term $\mathcal{T}^{(n)}$ are, respectively,  given by
\begin{eqnarray}
C_{\mu\nu}^{(n)}(q,r) & = & R_{n}(q_{\mu})R_{n}(r_{\nu})c_{\mu}(q)c_{\nu}(q)\Omega_{n}^{\mu\nu}(q+r),\nonumber\\
\mathcal{T}_{\mu\nu ; \rho\sigma}^{(n)} & =& \delta_{\rho\mu} \delta_{\sigma\mu}\mathcal{X}_{\mu\nu}^{(n)}(p,q) - \delta_{\rho\nu} \delta_{\sigma\nu}\mathcal{X}_{\nu\mu}^{(n)}(p,q) \nonumber\\
& & +\delta_{\rho\mu} \delta_{\sigma\nu}\mathcal{Y}_{\mu\nu}^{(n)}(p,q) - \delta_{\rho\nu} \delta_{\sigma\mu}\mathcal{Y}_{\nu\mu}^{(n)}(p,q), \nonumber
\end{eqnarray}
where
\begin{eqnarray}
\mathcal{X}_{\mu\nu}^{(n)}(p,q) & = & \frac{a}{2}\hat{q}_{\nu}R_{n}(q_{\mu}) R_{n}((p-q)_{\mu}) R_{n}(p_{\nu}),\nonumber\\
\mathcal{Y}_{\mu\nu}^{(n)}(p,q) & = & \frac{a}{2}\hat{q}_{\mu}R_{n}(q_{\mu}) R_{n}((p-q)_{\nu}) R_{n}(p_{\nu}).\nonumber
\end{eqnarray}
Thus
\begin{displaymath}
F_{\mu\nu}^{(2),a,3L} = \frac{1}{2}f^{abc}\int_{BZ}\frac{d^{4}q}{(2\pi)^4}\mathcal{S}_{\mu\nu;\rho\sigma}^{3L}(p,q;u_{0})A_{\rho}^{b}(q)A_{\sigma}^{c}(p-q),
\end{displaymath}
with
\begin{displaymath}
\mathcal{S}_{\mu\nu;\rho\sigma}^{(3L)}(p,q;u_{0})=\sum_{n=1}^{3}c_{n}^{(F)}\mathcal{S}_{\mu\nu;\rho\sigma}^{(n)}(p,q),
\end{displaymath}
and
\begin{displaymath}
c_1^{(F)}=\frac{3}{2u_0^4},\hspace{0.10cm}
c_2^{(F)}=-\frac{3}{20u_0^8},\hspace{0.10cm}
c_3^{(F)}=\frac{1}{90u_0^{12}}.
\end{displaymath}
Then the 3-gluon vertex is given by
\begin{eqnarray}
\Lambda_{\mu\nu;\rho\sigma\lambda}^{abc, (3)}(p,q,r; u_{0}) & =& \frac{2}{g_{0}}f^{abc}\left[\mathcal{G}_{\mu\nu;\rho\sigma\lambda}(p,q,r; u_{0}) \right. \nonumber\\
& & \left.  -\frac{1}{4}\delta_{\mu\nu}\mathcal{G}_{\alpha\alpha;\rho\sigma\lambda}(p,q,r; u_{0})\right],\nonumber\\
\label{eqV3}
\end{eqnarray}
where
\begin{eqnarray}
\mathcal{G}_{\mu\nu;\rho\sigma\lambda}(p,q,r; u_{0}) &=& \Omega_{3L}^{\mu\alpha}(p;u_{0})\big(\hat{p}_{\mu}\delta_{\alpha\rho}-\hat{p}_{\alpha}\delta_{\mu\rho}\big)\nonumber\\
& & \times \mathcal{S}_{\nu\alpha;\sigma\lambda}^{(3L)}(-p,q;u_{0}) \nonumber\\
& & + \Omega_{3L}^{\nu\alpha}(p;u_{0})\big(\hat{p}_{\nu}\delta_{\alpha\rho}-\hat{p}_{\alpha}\delta_{\nu\rho}\big)\nonumber\\
& & \times \mathcal{S}_{\mu\alpha;\sigma\lambda}^{(3L)}(-p,q;u_{0}),
\label{eqV3b}
\end{eqnarray}
with momentum conservation $p+q+r=0$.\\

The most involved ingredient is the four-gluon EMT vertex. The 4-gluon vertex 
\begin{eqnarray}
\Lambda_{\mu\nu;\rho\sigma\lambda\tau}^{abcd,(4)}(p_{i};u_{0}) &=& \Lambda_{\mu\nu;\rho\sigma\lambda\tau}^{abcd,(4)}|_{(2)(2)}(p_{i};u_{0}) \nonumber\\
& & +\Lambda_{\mu\nu;\rho\sigma\lambda\tau}^{abcd,(4)}|_{(1)(3)}(p_{i};u_{0})\nonumber
\end{eqnarray}
receives contributions from $F^{(2)}F^{(2)}$ and $F^{(1)}F^{(3)}$ parts. The  \(F^{(2)}F^{(2)}\) contribution can be written as
\begin{eqnarray}
\Gamma_{\mu\nu;\rho\sigma\lambda\tau}^{abcd,(4)}\Big|_{(2)(2)} &= &\sum_{\chi=s,t,u}\mathcal C_\chi^{abcd}
\left[ \mathcal H_{\mu\nu;\rho\sigma\lambda\tau}^{(\chi)}(p_i;u_0) \right. \nonumber\\
& & \left. -\frac14\delta_{\mu\nu} \mathcal H_{\alpha\alpha;\rho\sigma\lambda\tau}^{(\chi)}(p_i;u_0)\right],\nonumber
\label{eq:Gamma4_22_epjc}
\end{eqnarray}
with the color tensors
\begin{displaymath}
\mathcal C_s^{abcd}=f^{abe}f^{cde},\hspace{0.10cm}\mathcal C_t^{abcd}=f^{ace}f^{bde},\hspace{0.10cm}
\mathcal C_u^{abcd}=f^{ade}f^{bce}.
\label{eq:CsTu}
\end{displaymath}
Each \(\mathcal H^{(\chi)}\) is bilinear in \(\mathcal S^{(3L)}\), and therefore inherits the decomposition
\begin{equation}
\mathcal H = \mathcal H_{cc}
+\mathcal H_{cT}
+\mathcal H_{Tc}
+\mathcal H_{TT},
\label{eq:Hsplit}
\end{equation}
with $ \mathcal S=\mathcal S_c+\mathcal T$. Here $\mathcal H_{cc}$ is the core-core contribution, in which both quadratic field-strength factors are represented by their continuum-like local kernels. 
This term controls the leading continuum tensor structure of the operator. The mixed terms $\mathcal H_{cT}$ and $\mathcal H_{Tc}$ arise when one field-strength factor contributes through 
its core part and the other through its transport part, and therefore describe the leading interference between local continuum physics and the extended Wilson-loop geometry of the lattice clover operator. 
The transport-transport contribution $\mathcal H_{TT}$ is built solely from the transport kernels and represents the purely lattice-geometric, nonlocal correction. Thus, the decomposition provides 
a natural separation between continuum-like physics, leading lattice corrections, and fully geometric lattice artifacts.\\
For $s$ channel
\begin{eqnarray}
\mathcal H_{cc,\mu\nu;\rho\sigma\lambda\tau}^{s} &= & \frac{1}{4}\mathcal S^{(3L)}_{core,\mu\alpha;\rho\sigma}(p_{1}+p_{2},p_{1})\nonumber\\
& & \times \mathcal S^{(3L)}_{core,\nu\alpha;\lambda\tau}(p_{3}+p_{4},p_{3}+(\rho\sigma\leftrightarrow \lambda\tau ),\nonumber\\
\mathcal H_{cT,\mu\nu;\rho\sigma\lambda\tau}^{s}& = & \frac{1}{4}\mathcal S^{(3L)}_{core,\mu\alpha;\rho\sigma}(p_{1}+p_{2},p_{1})\nonumber\\
& & \times \mathcal T^{(3L)}_{core,\nu\alpha;\lambda\tau}(p_{3}+p_{4},p_{3}) +(\rho\sigma\leftrightarrow \lambda\tau ),\nonumber\\
\mathcal H_{Tc,\mu\nu;\rho\sigma\lambda\tau}^{s}& = & \frac{1}{4}\mathcal T^{(3L)}_{core,\mu\alpha;\rho\sigma}(p_{1}+p_{2},p_{1})\nonumber\\
& & \times \mathcal S^{(3L)}_{core,\nu\alpha;\lambda\tau}(p_{3}+p_{4},p_{3})+(\rho\sigma\leftrightarrow \lambda\tau ),\nonumber\\
\mathcal H_{TT,\mu\nu;\rho\sigma\lambda\tau}^{s}& = & \frac{1}{4}\mathcal T^{(3L)}_{core,\mu\alpha;\rho\sigma}(p_{1}+p_{2},p_{1})\nonumber\\
& & \times \mathcal T^{(3L)}_{core,\nu\alpha;\lambda\tau}(p_{3}+p_{4},p_{3})+(\rho\sigma\leftrightarrow \lambda\tau ).\nonumber
\end{eqnarray}
The \(F^{(1)}F^{(3)}\) contribution is written as:
\begin{eqnarray}
\Gamma_{\mu\nu;\rho\sigma\lambda\tau}^{abcd,(4)}\Big|_{(1)(3)} &=& 2\left[
\mathcal M_{\mu\nu;\rho\sigma\lambda\tau}^{abcd}(p_i;u_0)\right. \nonumber\\
& & \left. -\frac14\delta_{\mu\nu} \mathcal M_{\alpha\alpha;\rho\sigma\lambda\tau}^{abcd}(p_i;u_0)\right].
\label{eq:Gamma4_13_epjc}
\end{eqnarray}
For general kinematics the tensor \(\mathcal M\) is defined by a permutation sum over the four external gluon legs. 
However, in the operator tadpole diagram the kinematics specialize to \((p,-p,k,-k)\), and the permutation sum reduces to four non-equivalent assignments 
corresponding to the four choices of which external leg is absorbed by the linear kernel \(X\). In that case one may write the non-permutation form
\begin{eqnarray}
\mathcal M_{\mu\nu;\rho\sigma\alpha\beta}^{abcc}(p,-p,k,-k;u_0) =\nonumber\\
\hspace{2.0cm}X_{\mu\lambda;\rho}(p;u_0)\mathcal L_{\nu\lambda;\sigma\alpha\beta}^{\,bcc}(-p,-p,k;u_0)\nonumber\\
\hspace{2.0cm} +X_{\mu\lambda;\sigma}(-p;u_0)\mathcal L_{\nu\lambda;\rho\alpha\beta}^{\,acc}(p,k,-k;u_0)\nonumber\\
\hspace{2.0cm}+ X_{\mu\lambda;\alpha}(k;u_0)\mathcal L_{\nu\lambda;\rho\sigma\beta}^{\,abc}(-k,p,-p;u_0)\nonumber\\
\hspace{2.0cm}+X_{\mu\lambda;\beta}(-k;u_0)\mathcal L_{\nu\lambda;\rho\sigma\alpha}^{\,abc}(k,p,-p;u_0),\nonumber
\label{eq:M_noperm_epjc}
\end{eqnarray}
where
\begin{equation}
X_{\mu\lambda;\rho}(q;u_0)
=
\Omega_{3L}^{\mu\lambda}(q;u_0)
\big(\hat q_\mu\delta_{\lambda\rho}-\hat q_\lambda\delta_{\mu\rho}\big).
\label{eq:Xkernel_epjc}
\end{equation}
This representation is particularly convenient in the tadpole channel because it eliminates the explicit sum over \(S_4\) and makes the color contractions transparent.
The tadpole-specialized 4-gluon vertex, after color contraction of the two internal legs, becomes
\begin{eqnarray}
\Lambda_{\mu\nu;\rho\sigma\alpha\beta}^{abcd,(4)}(p,-p,k,-k; u_{0}) & = & C_{A}\delta^{ab}\left[\mathcal{S}_{\mu\lambda; \rho\alpha}(p,k; u_{0}) \right.\nonumber\\
& & \left. \times \mathcal{S}_{\mu\lambda; \sigma\beta}(-p,-k; u_{0}) \right.\nonumber\\
& & \left.  -\frac{1}{4}\delta_{\mu\nu} \mathcal{S}_{\eta\lambda; \rho\alpha}(p,k; u_{0})\right. \nonumber\\
& & \left. \times \mathcal{S}_{\eta\lambda; \sigma\beta}(-p,-k; u_{0})\right]\nonumber\\
& & +2C_{A}\delta^{ab}\left[\mathcal{M}_{\mu\nu;\rho\sigma\alpha\beta}(p,k; u_{0}) \right. \nonumber\\
& & \left. -\frac{1}{4}\delta_{\mu\nu}\mathcal{M}_{\lambda\lambda; \rho\sigma\alpha\beta}(p,k; u_{0})\right].\nonumber\\
\label{eq4V}
\end{eqnarray}
\subsection{One-loop Contributions}
\label{subsec:oneloop_emt}
The one-loop renormalization is extracted from the one-particle irreducible amputated two-gluon Green function with one EMT insertion,
\begin{equation}
\Gamma_{\mu\nu;\rho\sigma}^{ab}(p) = \left. \langle A_\rho^a(p)\,T_{\mu\nu}(0)\,A_\sigma^b(-p)\rangle \right|_{\mathrm{1PI,amp}}.
\label{eq:Lambda}
\end{equation}
At one loop, the correction separates into three topologies,
\begin{equation}
\Gamma^{(1)} = \Gamma_{\mathrm{sail}}^{(1)} +\Gamma_{\mathrm{tad}}^{(1)} +\Gamma_{\mathrm{leg}}^{(1)}.
\label{eq:one_loop_split}
\end{equation}
The  contributions to the one-loop correction, from  sail (external leg corrections),  operator self-energy (tadpole) diagrams, and  
vertex diagrams, are shown in Fig. \ref{fig:emt_oneloop_all}.  
The sail diagram is generated by the contraction of the three-gluon EMT vertex with the three-gluon gauge-action vertex through two internal gluon propagators.
These  represent the one-loop correction to the external gluon wave functions. The virtual gluon loop dresses the incoming or outgoing gluon before it interacts with the EMT operator.
These are wave function renormalization corrections to the external gluon legs.  These provide part of anomalous dimension (logarithmic divergence) and a lattice-specific finite contribution. 

\begin{figure}[h!]
\centering
\begin{tikzpicture}
\begin{feynman}
  \vertex (a) at (-2,0) {\(A_\rho^a(p)\)};
  \vertex[draw, rectangle, minimum width=1.2cm, minimum height=0.8cm] (op) at (0,0) {\(T_{\mu\nu}\)};
  \vertex (b) at (2,0) {\(A_\sigma^b(-p)\)};

  \diagram*{
    (a) -- [gluon, momentum=\(p\)] (op) -- [gluon, momentum=\(-p\)] (b),
  };
\end{feynman}
\end{tikzpicture}\\
\small (a) Tree-level operator insertion.

\vspace{0.4cm}

\centering
\begin{tikzpicture}
\begin{feynman}
  \vertex (a) at (-2,0) {\(A_\rho^a(p)\)};
  \vertex[draw, rectangle, minimum width=1.2cm, minimum height=0.8cm] (op) at (0,0) {\(T_{\mu\nu}\)};
  \vertex (v) at (2,0);
  \vertex (b) at (4,0) {\(A_\sigma^b(-p)\)};

  \diagram*{
    (a) -- [gluon, momentum=\(p\)] (op) -- [gluon, momentum=\(p+k\)] (v) -- [gluon, momentum=\(-p\)] (b),
    (op) -- [gluon, half left, looseness=1.2, momentum=\(k\)] (v),
    (v) -- [gluon, half left, looseness=1.2, momentum=\(-k-p\)] (op),
  };
\end{feynman}
\end{tikzpicture}\\
\small  (b) Sail.

\vspace{0.4cm}

\centering
\begin{tikzpicture}
\begin{feynman}
  \vertex (a) at (-2,0) {\(A_\rho^a(p)\)};
  \vertex[draw, rectangle, minimum width=1.2cm, minimum height=0.8cm] (op) at (0,0) {\(T_{\mu\nu}\)};
  \vertex (b) at (2,0) {\(A_\sigma^b(-p)\)};
  \vertex (u) at (0,1.4);

  \diagram*{
    (a) -- [gluon, momentum=\(p\)] (op) -- [gluon, momentum=\(-p\)] (b),
    (op) -- [gluon, half left, looseness=1.5, momentum=\(k\)] (u),
    (u) -- [gluon, half left, looseness=1.5, momentum=\(-k\)] (op),
  };
\end{feynman}
\end{tikzpicture}\\
\small  (c) Operator tadpole.

\vspace{0.4cm}

\centering
\begin{tikzpicture}
\begin{feynman}
  \vertex (a) at (-3,0) {\(A_\rho^a(p)\)};
  \vertex (s1) at (-1.5,0);
  \vertex[draw, rectangle, minimum width=1.2cm, minimum height=0.8cm] (op) at (1,0) {\(T_{\mu\nu}\)};
  \vertex (b) at (3,0) {\(A_\sigma^b(-p)\)};
  \vertex (u) at (-1.5,1.3);

  \diagram*{
    (a) -- [gluon, momentum=\(p\)] (s1) -- [gluon, momentum=\(p\)] (op) -- [gluon, momentum=\(-p\)] (b),
    (s1) -- [gluon, half left, looseness=1.4, momentum=\(k\)] (u),
    (u) -- [gluon, half left, looseness=1.6, momentum=\(k-p\)] (s1),
  };
\end{feynman}
\end{tikzpicture}\\
\small (d) Left external-leg self-energy correction.

\vspace{0.4cm}

\centering
\begin{tikzpicture}
\begin{feynman}
  \vertex (a) at (-3,0) {\(A_\rho^a(p)\)};
  \vertex[draw, rectangle, minimum width=1.2cm, minimum height=0.8cm] (op) at (-1,0) {\(T_{\mu\nu}\)};
  \vertex (s2) at (1.5,0);
  \vertex (b) at (3,0) {\(A_\sigma^b(-p)\)};
  \vertex (u) at (1.5,1.3);

  \diagram*{
    (a) -- [gluon, momentum=\(p\)] (op) -- [gluon, momentum=\(-p\)] (s2) -- [gluon, momentum=\(-p\)] (b),
    (s2) -- [gluon, half left, looseness=1.4, momentum=\(k\)] (u),
    (u) -- [gluon, half left, looseness=1.4, momentum=\(k+p\)] (s2),
  };
\end{feynman}
\end{tikzpicture}\\
\small  (e) Right external-leg self-energy correction.
\caption{
One-loop Feynman diagrams {\protect\footnotemark} contributing to the renormalization of the improved gluonic energy--momentum tensor $T_{\mu\nu}$.}
\label{fig:emt_oneloop_all}
\end{figure}
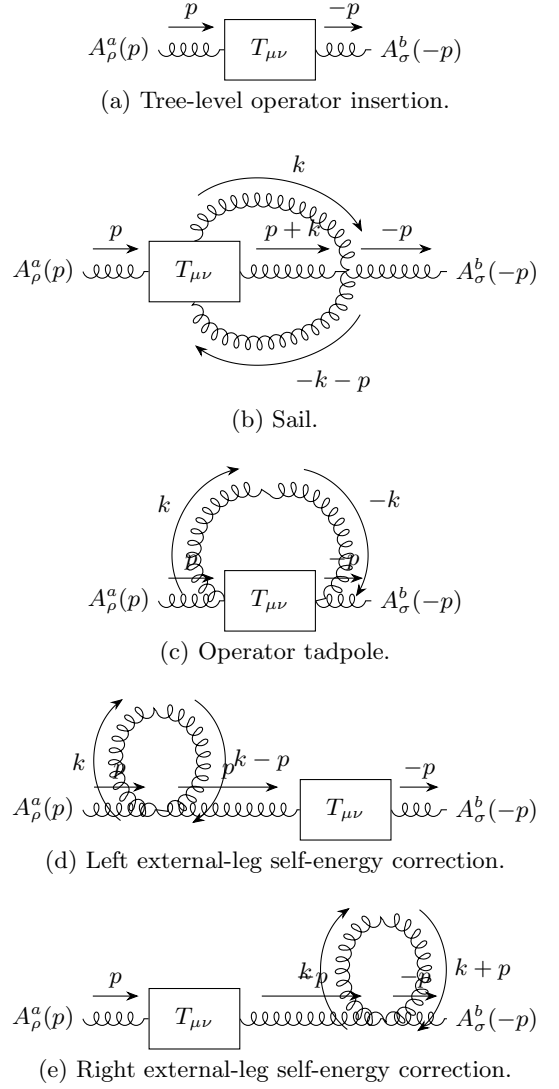
\footnotetext{The Feynman diagrams were generated using the LaTeX package \texttt{tikz-feynman}.}

In Landau gauge, the corresponding amputated two-gluon Green function is
\begin{eqnarray}
\Gamma_{\mu\nu;\rho\sigma}^{(1)\,\mathrm{sail},ab}(p) &
=&
\int_{BZ}\frac{d^4k}{(2\pi)^4}\,
\Lambda_{\mu\nu;\rho\alpha\beta}^{ace,(3)}(p,-p-k,k;u_0)\nonumber\\
& & \times D_{\alpha\alpha'}^{cc'}(p+k;u_0)\nonumber\\
& & \times V_{\sigma\alpha'\beta'}^{bc'e',(3)}(-p,p+k,-k;u_0) D_{\beta\beta'}^{ee'}(k;u_0), \nonumber\\
\label{eqsail}
\end{eqnarray}
where \(p\) is the external momentum and \(k\) is the loop momentum. The two internal propagators therefore carry momenta \(p+k\) and \(k\), respectively.
The 3-gluon gauge-action vertex for the tadpole-improved tree-level Symanzik action is
\begin{eqnarray}
V_{\sigma\alpha'\beta'}^{bc'e',(3)}(p_1,p_2,p_3;u_0) & =&g_0\,f^{bc'e'} \left[ \frac{c_0}{u_0^4}V_{\sigma\alpha'\beta'}^{P}(p_1,p_2,p_3) \right.\nonumber\\
& & \left. + \frac{c_1}{u_0^6}V_{\sigma\alpha'\beta'}^{R}(p_1,p_2,p_3)\right],
\label{eqsv}
\end{eqnarray}
where $V^P$ and $V^R$ denote the plaquette and rectangle parts, respectively.  For the sail kinematics used in Eq.~\eqref{eqsv},
\begin{equation}
(p_1,p_2,p_3)=(-p,p+k,-k).
\label{eq:sail_kinematics_epjc}
\end{equation}
It is convenient to define the transverse projector
\begin{equation}
P_{\alpha\beta}(q)\equiv
\delta_{\alpha\beta}-\frac{\hat q_\alpha\hat q_\beta}{\hat q^2},
\label{eq:sail_transverse_projector}
\end{equation}
so that
\begin{eqnarray}
D_{\alpha\alpha'}^{cc'}(p+k;u_0) &=&\delta^{cc'}\,\frac{P_{\alpha\alpha'}(p+k)}{\widehat{(p+k)}^2 K_T(p+k;u_0)},\nonumber\\
D_{\beta\beta'}^{ee'}(k;u_0) &=&\delta^{ee'}\,\frac{P_{\beta\beta'}(k)}{\hat k^2 K_T(k;u_0)},
\label{eqp}
\end{eqnarray}
with
\begin{displaymath}
K_T(q;u_0)=\tilde c_0+\tilde c_1 a^2\hat q^2, \hspace{0.10cm}
\tilde c_0=\frac{5}{3u_0^4}-\frac{2}{3u_0^6},
\hspace{0.10cm}
\tilde c_1=\frac{1}{12u_0^6},
\label{eq:sail_KT_epjc}
\end{displaymath}
and \(\hat q_\mu=(2/a)\sin(aq_\mu/2)\). 

Substituting Eqs.  (\ref{eqV3}), (\ref{eqsv}) and (\ref{eqp}) in Eq. (\ref{eqsail}), and noting that the explicit factors of \(g_0\) cancel, one obtains
\begin{eqnarray}
\Gamma_{\mu\nu;\rho\sigma}^{(1,\mathrm{sail},ab)}(p) & = & 2\int_{BZ}\frac{d^4k}{(2\pi)^4}f^{ace}\delta^{cc'}f^{bc'e'}\delta^{ee'}\nonumber\\
& &\frac{\mathcal N_{\mu\nu;\rho\sigma}^{\mathrm{sail}}(p,k;u_0)}{\widehat{(p+k)}^{2} K_{T}(p+k;u_0)\hat {k}^{2} K_{T}(k;u_0)},\nonumber
\label{eqsail2}
\end{eqnarray}
where the numerator is
\begin{eqnarray}
\mathcal N_{\mu\nu;\rho\sigma}^{\mathrm{sail}}(p,k;u_0)
&=&\left[\mathcal G_{\mu\nu;\rho\alpha\beta}(p,-p-k,k;u_0) \right. \nonumber\\
& & \left. -\frac14\delta_{\mu\nu}\mathcal G_{\lambda\lambda;\rho\alpha\beta}(p,-p-k,k;u_0)\right]\nonumber\\
& &\times  P_{\alpha\alpha'}(p+k)\left[\frac{c_0}{u_0^4}V_{\sigma\alpha'\beta'}^{P} \right.\nonumber\\
& & \left.+\frac{c_1}{u_0^6}V_{\sigma\alpha'\beta'}^{R}\right]P_{\beta\beta'}(k).
\label{eqsail3}
\end{eqnarray}
The color contraction is immediate:
\begin{equation}
f^{ace}\delta^{cc'}f^{bc'e'}\delta^{ee'}
=
f^{ace}f^{bce}
=
C_A\,\delta^{ab},
\label{eq:sail_color_epjc}
\end{equation}
and therefore the contribution from the sail diagram reduces to
\begin{eqnarray}
\Gamma_{\mu\nu;\rho\sigma}^{(1)\,\mathrm{sail},ab}(p) &=&2C_A\delta^{ab}\int_{BZ}\frac{d^4k}{(2\pi)^4}\nonumber\\
& & \times \frac{\mathcal N_{\mu\nu;\rho\sigma}^{\mathrm{sail}}(p,k;u_0)}{\widehat{(p+k)}^2 K_T(p+k;u_0)\,\hat k^2 K_T(k;u_0)}.\nonumber\\
\label{eqsail4}
\end{eqnarray}
Using Eq. (\ref{eqV3b}), the numerator may be written explicitly as:
\begin{eqnarray}
\mathcal N_{\mu\nu;\rho\sigma}^{\mathrm{sail}}(p,k;u_0) &= &\left[X_{\mu\lambda;\rho}(p;u_0)\mathcal S_{\nu\lambda ;\alpha\beta}^{(3L)}(-p,-p-k;u_0)\right.\nonumber\\
& & \left.  + X_{\nu\lambda;\rho}(p;u_0)\mathcal S_{\mu\lambda;\alpha\beta}^{(3L)}(-p,-p-k;u_0) \right. \nonumber\\
& & \left. -\frac{1}{2}\delta_{\mu\nu}X_{\eta\lambda;\rho}(p;u_0)\right. \nonumber\\
& & \left. \times \mathcal S_{\eta\lambda;\alpha\beta}^{(3L)}(-p,-p-k;u_0)\right]\nonumber\\
& & \times P_{\alpha\alpha'}(p+k) \left[ \frac{c_0}{u_0^4}V_{\sigma\alpha'\beta'}^{P}(-p,p+k,-k)\right. \nonumber\\
& & \left. + \frac{c_1}{u_0^6}V_{\sigma\alpha'\beta'}^{R}(-p,p+k,-k)\right]P_{\beta\beta'}(k).\nonumber\\
\label{eqs5}
\end{eqnarray}
It is useful to separate the core and transport pieces of the clover kernel:
\begin{displaymath}
\mathcal K^{(3L)}=\mathcal K_c^{(3L)}+\mathcal T^{(3L)}.
\end{displaymath}
Accordingly,
\begin{displaymath}
\mathcal N^{\mathrm{sail}} =\mathcal N_c^{\mathrm{sail}}+\mathcal N_T^{\mathrm{sail}},
\end{displaymath}
with
\begin{eqnarray}
\mathcal N_c^{\mathrm{sail}}(p,k;u_0) & =& \left[ X_{\mu\lambda;\rho}(p)\,\mathcal S_{c,\nu\lambda;\alpha\beta}^{(3L)}(-p,-p-k)\right.\nonumber\\
& & \left. + X_{\nu\lambda;\rho}(p)\,\mathcal S_{c,\mu\lambda;\alpha\beta}^{(3L)}(-p,-p-k) \right.\nonumber\\
& &  \left. -\frac{1}{2}\delta_{\mu\nu}X_{\eta\lambda;\rho}(p)\,\mathcal S_{c,\eta\lambda;\alpha\beta}^{(3L)}(-p,-p-k)\right]\nonumber\\
& &\times P_{\alpha\alpha'}(p+k)\left[\frac{c_0}{u_0^4}V_{\sigma\alpha'\beta'}^{P}\right.\nonumber\\
& & \left. +\frac{c_1}{u_0^6}V_{\sigma\alpha'\beta'}^{R}\right]P_{\beta\beta'}(k),
\label{eqN1}
\end{eqnarray}
and
\begin{eqnarray}
\mathcal N_T^{\mathrm{sail}}(p,k;u_0) & =&\left[ X_{\mu\lambda;\rho}(p)\mathcal T_{\nu\lambda;\alpha\beta}^{(3L)}(-p,-p-k)\right.\nonumber\\
& & \left. + X_{\nu\lambda;\rho}(p)\,\mathcal T_{\mu\lambda;\alpha\beta}^{(3L)}(-p,-p-k)\right. \nonumber\\
& & \left. -\frac12\delta_{\mu\nu}X_{\eta\lambda;\rho}(p)\,\mathcal T_{\eta\lambda;\alpha\beta}^{(3L)}(-p,-p-k)\right]\nonumber\\
& &\times P_{\alpha\alpha'}(p+k)\left[\frac{c_0}{u_0^4}V_{\sigma\alpha'\beta'}^{P}\right. \nonumber\\
& & \left. + \frac{c_1}{u_0^6}V_{\sigma\alpha'\beta'}^{R} \right] P_{\beta\beta'}(k).
\label{eqN2}
\end{eqnarray}
The first term \(\mathcal N_c^{\mathrm{sail}}\) contains the continuum-like vertex structure, while \(\mathcal N_T^{\mathrm{sail}}\) measures the leading distortion induced by the 
finite lattice geometry of the clover operator. This split is useful both conceptually and practically, since it separates the continuum matching structure from the genuinely lattice-specific correction.\\

To obtain the contribution relevant for the renormalization constant, the Green function is projected onto the transverse symmetric-traceless spin-2 channel. We use
\begin{eqnarray}
t_{\rho\sigma}(p) &=&\delta_{\rho\sigma}-\frac{\hat p_\rho\hat p_\sigma}{\hat p^2}, \nonumber\\
P_{\mu\nu;\gamma\delta} &=&\frac{1}{2}(\delta_{\mu\gamma}\delta_{\nu\delta}+\delta_{\mu\delta}\delta_{\nu\gamma}) -\frac14\delta_{\mu\nu}\delta_{\gamma\delta},\nonumber
\end{eqnarray}
and define
\begin{displaymath}
\mathcal F_{\mathrm{sail}}^{(1)}(p;u_0) = P_{\mu\nu;\gamma\delta} t_{\rho\gamma}(p)t_{\sigma\delta}(p)\Gamma_{\mu\nu;\rho\sigma}^{(1)\mathrm{sail}}(p;u_0).
\end{displaymath}
Substituting Eq. (\ref{eqsail4}) into the above equation gives
\begin{eqnarray}
\mathcal F_{\mathrm{sail}}^{(1)}(p;u_0) &=&2C_A \int_{BZ}\frac{d^4k}{(2\pi)^4}\nonumber\\
& & \times  \frac{P_{\mu\nu;\gamma\delta}t_{\rho\gamma}(p)t_{\sigma\delta}(p)\mathcal N_{\mu\nu;\rho\sigma}^{\mathrm{sail}}(p,k;u_0)}{\widehat{(p+k)}^2 K_T(p+k;u_0)\,\hat k^2 K_T(k;u_0)}.\nonumber\\
\end{eqnarray}
Expanding the projected integrand for small external momentum and retaining the coefficient of the \(p^2\) tensor structure, the sail diagram contributes to the finite one-loop matching coefficient through
\begin{equation}
\mathcal F_{\mathrm{sail}}^{(1)}(p;u_0)=\frac{g_0^2C_A}{16\pi^2}\,p^2\,B_{\mathrm{sail}}(u_0)+O(p^4),
\label{eq:sail_Bdef_epjc}
\end{equation}
where $B_{\mathrm{sail}}(u_0)$  is reduced analytically to the minimal basis of Brillouin-zone integrals 
\begin{eqnarray}
B_{sail}(u_{0}) & = &\frac{5}{3}\tilde{c}_{0}^{2}I_{0}(u_{0}) - \frac{1}{3}\tilde{c}_{0}^{2}I_{1}(u_{0}) +\tilde{c}_{0}^{2}\tilde{c}_{1}I_{2}(u_{0})\nonumber\\
& & - \frac{7}{6}\tilde{c}_{0}^{2}\sum_{n,m=1}^{3}c_{n}^{(F)}c_{m}^{(F)}I_{nm}^{(0)}(u_{0})\nonumber\\
& & + \frac{1}{2}\tilde{c}_{0}^{2}\tilde{c}_{1}\sum_{n,m=1}^{3}c_{n}^{(F)}c_{m}^{(F)}I_{nm}^{(1)}(u_{0}).
\end{eqnarray}
For the isotropic kernel \(K_T(k;u_0)=\tilde c_0+\tilde c_1 a^2\hat k^2\), the one-loop projected amplitude reduces to the minimal scalar basis
\begin{eqnarray}
I_0(u_0) &= & \int_{BZ}\frac{d^4k}{(2\pi)^4}\frac{1}{\hat k^2 K_T(k;u_0)^2}, \nonumber\\
I_1(u_0) &= &\int_{BZ}\frac{d^4k}{(2\pi)^4}\frac{1}{K_T(k;u_0)^2},\nonumber\\
I_2(u_0) &=&\int_{BZ}\frac{d^4k}{(2\pi)^4}\frac{1}{\hat k^2 K_T(k;u_0)^3}, \nonumber
\end{eqnarray}
together with the operator-weighted integrals
\begin{eqnarray}
I_{nm}^{(0)}(u_0) &=&\int_{BZ}\frac{d^4k}{(2\pi)^4}\frac{\Phi_{nm}(k)}{\hat k^2 K_T(k;u_0)^2},\nonumber\\
I_{nm}^{(1)}(u_0) &=&\int_{BZ}\frac{d^4k}{(2\pi)^4}\frac{\Phi_{nm}(k)}{\hat k^2 K_T(k;u_0)^3},\nonumber
\end{eqnarray}
where
\begin{equation}
\Phi_{nm}(k)=\frac14\sum_{\mu<\nu}\Omega_n^{\mu\nu}(k)\Omega_m^{\mu\nu}(k).
\label{eq:Phi_nm_epjc}
\end{equation}
In this way the sail contribution is represented by an explicit, gauge-invariant numerator built from the improved EMT kernel, the improved gauge-action vertex, and 
the transverse lattice propagator, together with a scalar denominator controlled entirely by the improved transverse kernel \(K_T(k;u_0)\). \\

The operator tadpole diagrams  arise when the two gluon fields in the operator's $O(A^{2})$ vertex  contract with each other, forming a closed loop. 
The operator "feeds on itself" by emitting and reabsorbing virtual gluons. These  diagrams are expected to give power divergences that must be subtracted.
The contribution from the operator tadpole is given by
\begin{eqnarray}
\Gamma_{\mu\nu;\rho\sigma}^{(1)\,\mathrm{tad},ab}(p) &=& \frac12\int_{BZ}\frac{d^4k}{(2\pi)^4}\Lambda_{\mu\nu;\rho\sigma\alpha\beta}^{abcc,(4)}(p,-p,k,-k;u_0)\nonumber\\
& & \times D_{\alpha\beta}^{cc}(k;u_0).\nonumber
\label{eqtad1}
\end{eqnarray}
Using the propagator, the above expression becomes
\begin{equation}
\Gamma_{\mu\nu;\rho\sigma}^{(1)\,\mathrm{tad},ab}(p) =g_{0}^{2}C_{A}\delta^{ab}\int_{BZ}\frac{d^4k}{(2\pi)^4}\frac{\mathcal N_{\mu\nu;\rho\sigma}^{tad}(p,k;u_0)}{\hat{k}^{2}K_{T}(k;u_{0})}
\label{eqtad2}
\end{equation}
with
\begin{displaymath}
\mathcal N_{\mu\nu;\rho\sigma}^{tad} = \frac{1}{2}\Lambda_{\mu\nu;\rho\sigma\alpha\beta}^{abcc,(4)}(p,-p,k,-k;u_0)\bigg( \delta_{\alpha\beta}-\hat{k}_{\alpha}\hat{k}_{\beta}/\hat{k}^{2}\bigg)
\end{displaymath}

The vertex diagrams involve the interaction between the operator and the action vertices. The operator vertex (from expanding $T_{\mu\nu}^{Imp}$ to $O(A^{2})$) connects 
with the action vertex (three-gluon vertex) via gluon propagators. The external-leg contribution  generated by the gluon self-energy, is given by
\begin{equation}
\Pi_{\rho\sigma}^{ab}(p)
=
\delta^{ab}
\left[
(\delta_{\rho\sigma}\hat p^2-\hat p_\rho\hat p_\sigma)\Pi_T(p^2;u_0)
+\hat p_\rho\hat p_\sigma\Pi_L(p^2;u_0)
\right],
\label{eq:self_epjc}
\end{equation}
and contributes through
\begin{equation}
\Gamma_{\mu\nu;\rho\sigma}^{(1)\,\mathrm{leg}}(p)=-2g_0^2\,\Pi_T(\mu^2;u_0)\,\Lambda_{\mu\nu;\rho\sigma}^{(2)}(p;u_0).
\label{eqleg}
\end{equation}
To extract the renormalization constant, we project onto the transverse symmetric-traceless channel 
and define
\begin{equation}
\mathcal F^{\mathrm{lat}}(p;u_0)=P_{\mu\nu;\alpha\beta}\, t_{\rho\alpha}(p)t_{\sigma\beta}(p)\,
\Lambda_{\mu\nu;\rho\sigma}^{\mathrm{lat}}(p;u_0).
\label{eqFproj}
\end{equation}
The total coefficient is the sum of three one-loop topologies
\begin{displaymath}
\mathfrak B_{\mathrm{lat}}(u_0) = \mathfrak B_{\mathrm{lat}}^{\mathrm{sail}}(u_0)+\mathfrak B_{\mathrm{lat}}^{\mathrm{tad}}(u_0)+\mathfrak B_{\mathrm{lat}}^{\mathrm{leg}}(u_0)
\end{displaymath}
where
\begin{eqnarray}
\mathfrak B_{\mathrm{lat}}^{\mathrm{sail}}(u_0) &=& C_0^{\mathrm{sail}}(u_0)I_0(u_0)+C_1^{\mathrm{sail}}(u_0)I_1(u_0)\nonumber\\
& & +C_2^{\mathrm{sail}}(u_0)I_2(u_0)\nonumber\\
& & +\sum_{n,m=1}^3\left[C_{nm}^{(0),\mathrm{sail}}(u_0)I_{nm}^{(0)}(u_0)\right.\nonumber\\
& & \left. +C_{nm}^{(1),\mathrm{sail}}(u_0)I_{nm}^{(1)}(u_0)\right],\nonumber\\
\mathfrak B_{\mathrm{lat}}^{\mathrm{tad}}(u_0) &=& C_0^{\mathrm{tad}}(u_0)I_0(u_0)+\sum_{n,m=1}^3 C_{nm}^{(0),\mathrm{tad}}(u_0)I_{nm}^{(0)}(u_0),\nonumber\\
\mathfrak B_{\mathrm{lat}}^{\mathrm{leg}}(u_0) &=& C_0^{\mathrm{leg}}(u_0)I_0(u_0)+C_1^{\mathrm{leg}}(u_0)I_1(u_0)\nonumber\\
& & +C_2^{\mathrm{leg}}(u_0)I_2(u_0), \nonumber
\end{eqnarray}
with
\begin{eqnarray}
C_0^{\mathrm{sail}}(u_0)&=& \frac53\,\tilde c_0(u_0)^2, \hspace{0.10cm} C_1^{\mathrm{sail}}(u_0)=-\frac13\tilde c_0(u_0)^2,\nonumber\\
C_2^{\mathrm{sail}}(u_0)&=&\tilde c_0(u_0)^2\tilde c_1(u_0),\nonumber\\
C_{nm}^{(0),\mathrm{sail}}(u_0)&=&-\frac76 c_n^{(F)}c_m^{(F)}\tilde c_0(u_0)^2,\nonumber\\
C_{nm}^{(1),\mathrm{sail}}(u_0)&=&\frac12 c_n^{(F)}c_m^{(F)}\tilde c_0(u_0)^2\tilde c_1(u_0),\nonumber\\
C_0^{\mathrm{tad}}(u_0)&=&-\frac14 \tilde c_0(u_0) \mathcal N_F(u_0),\nonumber\\
C_{nm}^{(0),\mathrm{tad}}(u_0)&=&\frac32 c_n^{(F)}c_m^{(F)}\tilde c_0(u_0),\nonumber\\
C_0^{\mathrm{leg}}(u_0)&=&\frac{13}{3}\mathcal N_F(u_0)\tilde c_0(u_0)^2,\nonumber\\
C_1^{\mathrm{leg}}(u_0)&=&-\frac{2}{3}\mathcal N_F(u_0)\tilde c_0(u_0)^2,\nonumber\\
C_2^{\mathrm{leg}}(u_0)&=&2\,\mathcal N_F(u_0)\tilde c_0(u_0)^2\tilde c_1(u_0),\nonumber
\end{eqnarray}
and
\begin{displaymath}
\mathcal N_F(u_0) = 6\bigg(\frac{3}{2u_{0}^{4}} - \frac{3}{5u_{0}^{8}} +\frac{1}{10u_{0}^{12}}\bigg)^{2}.
\end{displaymath}
The renormalization constant is then given by
\begin{equation}
Z_{T}(u_{0})    = 1+\frac{g_0^2C_A}{16\pi^2}\left[\mathfrak B_{\overline{\mathrm{MS}}} -\mathfrak B_{\mathrm{lat}}(u_0)\right]+O(g_0^4).
\end{equation}
The renormalized traceless EMT obtained from the one-loop matching is finally defined by
\begin{equation}
T_{\mu\nu}^{R,\mathrm{TL}} =\left[1+\frac{g_0^2C_A}{16\pi^2}\bigg(\mathfrak B_{\overline{\mathrm{MS}}} -\mathfrak B_{\mathrm{lat}}(u_0)\bigg)\right]T_{\mu\nu}^{\mathrm{TL}}+O(g_{0}^{4}).
\label{eqZT}
\end{equation}

\subsection{Yang-Mill Trace Anamoly}
The one-loop renormalization derived in the spin-2 channel determines the normalization of the \emph{traceless} part of the improved gluonic energy--momentum tensor (EMT), 
but it does not by itself account for the anomalous trace. For the tadpole-improved operator,
\begin{equation}
T_{\mu\nu}^{\mathrm{imp}}(x)
=
\frac{1}{g_0^2}
\left[
F_{\mu\alpha}^{3L}(x)F_{\nu\alpha}^{3L}(x)
-\frac14\delta_{\mu\nu}\,
F_{\alpha\beta}^{3L}(x)F_{\alpha\beta}^{3L}(x)
\right],
\label{eq:Timptrace}
\end{equation}
the classical trace vanishes identically in \(d=4\),
\begin{equation}
\delta_{\mu\nu}T_{\mu\nu}^{\mathrm{imp}}(x)=0,
\label{eq:classicaltracezero}
\end{equation}
so the operator is classically traceless. The quantum trace anomaly therefore cannot originate from the multiplicative spin-2 renormalization constant alone. Instead, 
it resides in the scalar trace channel and must be incorporated through the full renormalized EMT rather than through its traceless projection only.
Accordingly, the renormalized EMT must therefore be written as
\begin{equation}
T_{\mu\nu}^{R}
=
Z_T(u_0)\,T_{\mu\nu}^{\mathrm{imp,TL}}
+
\frac14\delta_{\mu\nu}\,
Z_{\Theta}(u_0,a\mu)\,
\Theta^{\mathrm{imp}}(a),
\label{eq:fullEMTlattice}
\end{equation}
with
\begin{equation}
\Theta^{\mathrm{imp}}(a)
=
\frac{1}{g_0^2}\,\mathcal O_S^{\mathrm{imp}}(a)
=
\frac{1}{g_0^2}\,
F_{\rho\sigma}^{3L}(a)F_{\rho\sigma}^{3L}(a).
\label{eq:Thetaimp}
\end{equation}
The anomaly condition requires
\begin{equation}
\Theta^{R}(\mu)
=
\frac{\beta(g)}{2g}\,
F_{\rho\sigma}^{a}F_{\rho\sigma}^{a},
\label{eq:anomalycondition}
\end{equation}
or equivalently
\begin{equation}
Z_{\Theta}(u_0,a\mu)\,
\frac{1}{g_0^2}\,
F_{\rho\sigma}^{3L}F_{\rho\sigma}^{3L}
=
\frac{\beta(g)}{2g}\,
F_{\rho\sigma}^{a}F_{\rho\sigma}^{a}.
\label{eq:ZThetamatching}
\end{equation}
Since the improved clover combination reproduces the continuum field strength up to discretization errors,
\begin{equation}
F_{\rho\sigma}^{3L}F_{\rho\sigma}^{3L}
=
F_{\rho\sigma}^{a}F_{\rho\sigma}^{a}
+O(a^2),
\label{eq:F3Lcontinuumlimit}
\end{equation}
the matching condition becomes
\begin{equation}
\frac{Z_{\Theta}(u_0,a\mu)}{g_0^2}
=
\frac{\beta(g)}{2g}
+O(g^4,a^2).
\label{eq:ZThetafinalmatch}
\end{equation}
Thus the \(u_0\)-improved construction modifies the finite lattice matching coefficients, but it does not alter the anomaly coefficient itself.

At one loop in pure Yang--Mills theory,
\begin{equation}
\beta(g)
=
-\beta_0\,\frac{g^3}{16\pi^2}
+O(g^5),
\qquad
\beta_0=\frac{11}{3}C_A,
\label{eq:beta0pureYM}
\end{equation}
so that
\begin{equation}
\frac{\beta(g)}{2g}
=
-\frac{\beta_0}{2}\frac{g^2}{16\pi^2}
+O(g^4)
=
-\frac{11\,C_A}{6}\frac{g^2}{16\pi^2}
+O(g^4).
\label{eq:betaover2g}
\end{equation}
Therefore the renormalized trace must be
\begin{equation}
T^{\mu}_{\ \mu}
=
-\frac{11\,C_A}{6}\,
\frac{g^2}{16\pi^2}\,
F_{\rho\sigma}^{a}F_{\rho\sigma}^{a}
+O(g^4),
\label{eq:onelooptraceanomaly}
\end{equation}
which is precisely the one-loop Yang--Mills trace anomaly. In the improved lattice formulation this implies
\begin{equation}
T^{\mu}_{\ \mu,R}
=
-\frac{11\,C_A}{6}\,
\frac{g^2}{16\pi^2}\,
F_{\rho\sigma}^{a}F_{\rho\sigma}^{a}
+O(g^4,a^2).
\label{eq:latticetraceoneloop}
\end{equation}

An equivalent lattice interpretation follows from the relation between the EMT trace and the response of the action to a scale variation or anisotropy. 
Since \(T^{\mu}_{\ \mu}\sim a\,\partial S/\partial a\), while the bare action depends on \(a\) through the bare coupling \(g_0(a)\), one finds
\begin{equation}
a\frac{\partial S}{\partial a}
=
\frac{\partial S}{\partial g_0}\,
a\frac{dg_0}{da}
\propto
\beta(g)\,\frac{\partial S}{\partial g},
\label{eq:latticebetafunctionroute}
\end{equation}
and, because \(\partial S/\partial g\propto g^{-3}F^2\), this again yields
\begin{equation}
T^{\mu}_{\ \mu}
=
\frac{\beta(g)}{2g}\,F^2.
\label{eq:anisotropytrace}
\end{equation}
Hence the anomaly follows directly from the running of the coupling in the gauge action, and the improved EMT must reproduce it because it is 
built from the same improved field-strength basis that matches the continuum action density.

Combining the traceless and scalar sectors, the correct renormalized EMT is therefore
\begin{equation}
T_{\mu\nu}^{R}
=
Z_T(u_0)\,T_{\mu\nu}^{\mathrm{imp,TL}}
+
\frac14\delta_{\mu\nu}
\left[
\frac{\beta(g)}{2g}\,
F_{\rho\sigma}^{a}F_{\rho\sigma}^{a}
\right]
+O(a^2,g^4).
\label{eq:fullrenormEMTtrace}
\end{equation}
Taking the trace gives
\begin{equation}
\delta_{\mu\nu}T_{\mu\nu}^{R}
=
\frac{\beta(g)}{2g}\,
F_{\rho\sigma}^{a}F_{\rho\sigma}^{a}
+O(a^2,g^4),
\label{eq:finaltracecheck}
\end{equation}
because the traceless component satisfies
\begin{equation}
\delta_{\mu\nu}T_{\mu\nu}^{R,\mathrm{TL}}=0.
\label{eq:TLdropsout}
\end{equation}
Thus, the renormalization procedure based on the improved clover EMT correctly reproduces the quantum trace anomaly while 
preserving the properly normalized spin-2 part of the gluonic energy-momentum tensor.

We define the connected renormalized Euclidean correlator in the $00$ channel by
\begin{eqnarray}
G_{00}^{R}(\tau) &\equiv & \int d^3x\, \Big\langle \delta T_{00}^{R}(\tau,\mathbf{x})\, \delta T_{00}^{R}(0,\mathbf{0}) \Big\rangle, \nonumber\\
\delta T_{00}^{R} & \equiv &  T_{00}^{R}-\langle T_{00}^{R}\rangle . \nonumber
\end{eqnarray}
The corresponding dimensionless normalized correlator is
\begin{equation}
C_{00}^{R}(\tau)
\equiv
\frac{1}{T^5}\,G_{00}^{R}(\tau)
=
\frac{1}{T^5}
\int d^3x\,
\Big\langle
\delta T_{00}^{R}(\tau,\mathbf{x})\,
\delta T_{00}^{R}(0,\mathbf{0})
\Big\rangle.
\end{equation}
Using
\begin{eqnarray}
T_{\mu\nu}^{R} & =& Z_T(u_0)\,T_{\mu\nu}^{\mathrm{imp,TL}} + \frac14\,\delta_{\mu\nu}\,\Theta +O(a^2,g^4),\nonumber\\
\Theta &\equiv & \frac{\beta(g)}{2g}F^a_{\rho\sigma}F^a_{\rho\sigma},
\end{eqnarray}
the $00$ component becomes
\begin{equation}
T_{00}^{R} = Z_T(u_0)\,T_{00}^{\mathrm{imp,TL}} + \frac14\,\Theta +O(a^2,g^4),
\end{equation}
and hence
\begin{eqnarray}
G_{00}^{R}(\tau) & = & Z_T^2(u_0)\,G_{00}^{\mathrm{imp,TL}}(\tau) + \frac{Z_T(u_0)}{2}\,G_{0\Theta}^{\mathrm{mix}}(\tau)\nonumber\\
& & + \frac1{16}\,G_{\Theta\Theta}(\tau) +O(a^2,g^4),
\end{eqnarray}
where
\begin{eqnarray}
G_{00}^{\mathrm{imp,TL}}(\tau) & \equiv & \int d^{3}x\Big\langle\delta T_{00}^{\mathrm{imp,TL}}(\tau,\mathbf{x})\delta T_{00}^{\mathrm{imp,TL}}(0,\mathbf{0}) \Big\rangle\nonumber\\
G_{0\Theta}^{\mathrm{mix}}(\tau) & \equiv & \int d^{3}x \Big\langle \delta T_{00}^{\mathrm{imp,TL}}(\tau,\mathbf{x})\delta\Theta(0,\mathbf{0}) \nonumber\\
 & & + \delta\Theta(\tau,\mathbf{x})\,\delta T_{00}^{\mathrm{imp,TL}}(0,\mathbf{0}) \Big\rangle,\nonumber\\
G_{\Theta\Theta}(\tau) & \equiv &  \int d^{3}x\Big\langle \delta\Theta(\tau,\mathbf{x})\,\delta\Theta(0,\mathbf{0}) \Big\rangle.\nonumber
\end{eqnarray}
Therefore,
\begin{eqnarray}
C_{00}^{R}(\tau) & = & \frac{1}{T^5} \left[Z_T^2(u_0)\,G_{00}^{\mathrm{imp,TL}}(\tau) + \frac{Z_T(u_0)}{2}\,G_{0\Theta}^{\mathrm{mix}}(\tau) \right. \nonumber\\
& & \left.  + \frac1{16}\,G_{\Theta\Theta}(\tau) \right] +O(a^2,g^4).
\end{eqnarray}
In our formulation, the renormalized energy density $T_{00}^{R}$ receives both a multiplicative factor $Z_{T}(u_{0})$ acting on 
the improved traceless operator and an additive contribution from the trace-anomaly operator $\Theta = [\beta (g)/2g]F^{2}$.
Consequently, the correlator $C_{00}^{R}(\tau )$ is not simply a rescaled version of the bare correlator, but contains additional mixed and pure-anomaly contributions. 
At leading order the  renormalization affects both the overall magnitude and the detailed structure of the correlator. 
Physically, this modifies the short-distance behavior and can introduce deviations from an 
ideal plateau at intermediate $\tau T$ due to mixing with the scalar (trace) channel. 
Proper renormalization is therefore essential to ensure that the extracted plateau value correctly reproduces 
the thermodynamic observable $c_{V}/T^{3}$. 


\section{Comparison of one-loop perturbation theory with the lattice data}
\label{subsec:ZT_validation}
To assess the reliability of the perturbative matching factor \(Z_T(u_0)\), it is essential to compare the one-loop prediction with nonperturbative lattice data obtained 
from the same improved gauge action and operator basis to quantify the size of residual higher-order and discretization effects,  thereby indicating the range of lattice spacings for which the one-loop expression remains quantitatively reliable.
The inferred continuum finite part \(\mathfrak{B}_{\overline{\mathrm{MS}}}^{\mathrm{eff}}\) was determined by matching the lattice-side one-loop formula to the published dashed perturbative curve for \(Z_T(g_0^2)\) \cite{Giusti2015};
\begin{displaymath}
\mathfrak{B}_{\overline{\mathrm{MS}}}^{\mathrm{eff}} = \mathfrak{B}_{\mathrm{lat}}(u_{0})+ \frac{16\pi^2}{g_0^2 C_A}\bigl[Z_T^{\rm curve}(g_0^2)-1\bigr].
\end{displaymath}
Applying this point by point to the available \((g_0^2,u_0)\) values yielded a nearly constant result, from which the effective value $\mathfrak{B}_{\overline{\mathrm{MS}}}^{\mathrm{eff}}=15.04$ was extracted and is approximately constant across the relevant coupling range.
Thus \(15.04\) should be understood as the effective continuum finite matching coefficient that reproduces the published perturbative one-loop normalization in the same projector convention, rather than as a universal constant of the theory.
Thus the one-loop renormalization factor is computed as
\begin{equation}
Z_T^{\rm 1\mbox{-}loop}(u_0,g_0^2)=1+\frac{g_0^2 C_A}{16\pi^2}\bigl[15.04 - \mathfrak{B}_{\mathrm{lat}}(u_{0})\bigr].
\label{eqZ1-loop}
\end{equation}
By contrast, the pointwise values $\mathfrak{B}_{\overline{\mathrm{MS}}}^{\mathrm{eff}}(g_{0}^{2})$ extracted from non-perturbative lattice data should not be used as input for a 
one-loop prediction, since they already absorb higher-order and non-perturbative effects.
\begin{table*}[ht]
\centering
\caption{One-loop lattice contributions and renormalization factor computed with the effective inferred continuum finite term \(\mathfrak{B}_{\overline{\mathrm{MS}}}^{\mathrm{eff}}\).}
\label{tab:ZT_oneloop}
\begin{tabular}{cccccccc}
\hline
\(g_0^2\) & \(u_0\) & \(\mathfrak{B}_{\mathrm{sail}}\) &  \(\mathfrak{B}_{\mathrm{tad}}\) & \(\mathfrak{B}_{\mathrm{leg}}\)   &  \(\mathfrak{B}_{\mathrm{lat}}\) &\(Z_T^{\mathrm{1\mbox{-}loop}}\) \\
\hline
0.335 & 0.952 & -0.204527 & 0.131404 & 0.484903 & 0.411779 & 1.093(2) \\
0.442 & 0.943 & -0.209848 & 0.133627 & 0.505208 & 0.428987 & 1.123(2) \\
0.512 & 0.937 & -0.213314 & 0.135036 & 0.518870 & 1 0.440591 & 1.149(3) \\
0.605 & 0.930 & -0.217255 & 0.136592 & 0.534907  & 0.454244 & 1.168(3) \\
0.750 & 0.901 & -0.231816 & 0.141737 & 0.601840  & 0.511760 & 1.207(7) \\
0.850 & 0.872 & -0.241867 & 0.143884 & 0.668342  & 0.570359 & 1.236(7) \\
0.930 & 0.861 & -0.243974 & 0.143618 & 0.693388  & 0.593032 & 1.255(7) \\
1.000 & 0.853 & -0.244781 & 0.142970 & 0.711654  & 0.609842 & 1.274(6) \\
\hline
\end{tabular}
\end{table*}
Table~\ref{tab:ZT_oneloop} shows that the dominant contribution to \(\mathfrak{B}_{\mathrm{lat}}(u_0)\) comes from the external-leg term, while the tadpole contribution is 
positive but much smaller, and the sail contribution is negative. The corrected one-loop prediction \(Z_T^{\mathrm{1\mbox{-}loop}}\) follows the perturbative curve closely but remains systematically below the continuum-extrapolated lattice data, 
with the discrepancy increasing monotonically from weak to moderate coupling. The discrepancy grows with \(g_0^2\): it is  
moderate at intermediate coupling, and becomes large near \(g_0^2\simeq 1\), where the lattice value is about \(1.6\) while the one-loop curve is only about \(1.27\). 
This indicates that the one-loop calculation captures the leading trend but misses sizable higher-order and/or nonperturbative contributions, which enhance the growth of \(Z_T\) at moderate coupling.
This provides quantitative evidence that the one-loop matching captures the correct normalization trend but does not fully account for the stronger growth of \(Z_T\) seen in the lattice calculation. 
This strongly suggests that  higher-order perturbative terms, and possibly additional nonperturbative effects, are required for quantitative agreement beyond the weak-coupling regime\\

To make the comparison with Fig. \ref{fig4} more quantitative, we fit the continuum-extrapolated lattice data for \(Z_T\) as a function of \(g_0^2\) with a simple cubic ansatz,
\begin{equation}
Z_T^{\rm lat}(g_0^2) = a_0+a_1 g_0^2+a_2 (g_0^2)^2+a_3 (g_0^2)^3.
\label{eq:ZT_cubic}
\end{equation}
and obtain the effective two-loop and cubic fits using the lattice data from Ref. \cite{Giusti2015} to provide a phenomenological estimate of higher-loop effects beyond the one-loop calculation.
These fits reproduce the lattice curve accurately over the interval \(0\le g_0^2\le 1\), with an rms deviation of order \(6\times10^{-3}\).\\
In our explicit calculation, however, the perturbative curve is understood as the full matching factor.
The lattice  finite contribution itself is comparatively small, \(\mathfrak{B}_{\mathrm{lat}}(u_0)\simeq 0.41\text{--}0.61\), so the magnitude of the perturbative curve is 
controlled predominantly by the continuum--lattice difference \(\mathfrak{B}_{\overline{\mathrm{MS}}}^{\mathrm{eff}}-\mathfrak{B}_{\mathrm{lat}}(u_0)\).
 \begin{figure}[h!]
\scalebox{0.70}{\includegraphics{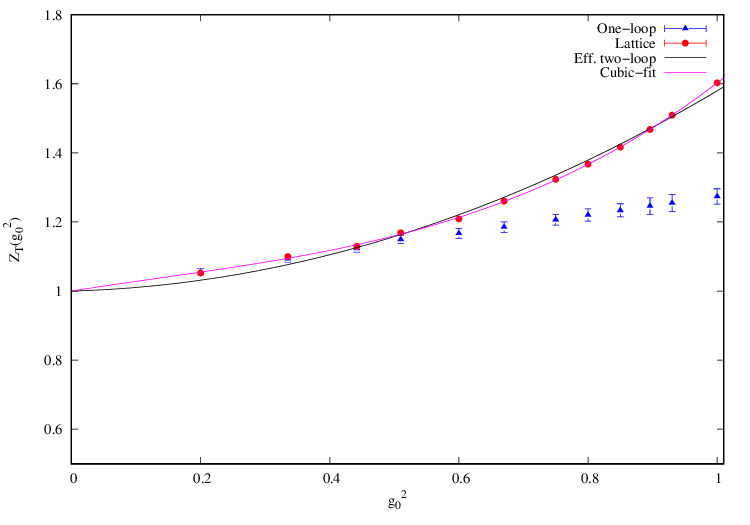}} 
\caption{\label{fig4} The renormalization factor $Z_{T}(g_{0}^{2})$  together with  the one-loop analytic result in Eq. (\ref{eqZ1-loop})}
\end{figure}
The comparison between Eqs.~\eqref{eqZ1-loop}  and \eqref{eq:ZT_cubic} and makes the origin of the discrepancy transparent. 
\noindent Because the present operator is tadpole improved and built from the three-loop clover combination, one expects the one-loop prediction to show 
significantly improved agreement with lattice data compared with the unimproved plaquette-clover operator. In particular, the explicit \(u_0\)-dependence resums 
an important subset of mean-field corrections and should reduce the magnitude of finite renormalization effects at moderate lattice couplings.\\

 \begin{figure}[h!]
\scalebox{0.70}{\includegraphics{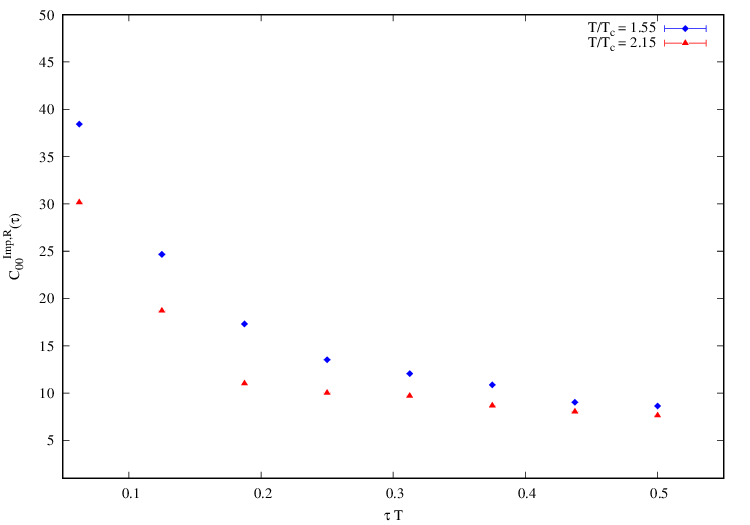}} 
\caption{\label{fig6} The anomaly correlator $C_{00}(\tau T)$  on $N_{\tau}=16$ lattices. }
\end{figure}
In figure \ref{fig6} we plot the normalized energy-density correlator $C_{00}(\tau T)$ for $T/T_{c}=1.55$ and $2.15$. The correlator shows a pronounced decrease 
with increasing time separation, with the strongest signal at small $\tau T$ and a gradual flattening toward the midpoint $\tau T=0.5$. Over the full range, the 
magnitude of the 00-channel fluctuations increases with temperature. The enhancement at small $\tau T$ is consistent with dominant short-distance ultraviolet 
contributions, while the weaker variation at larger $\tau T$ indicates that these ultraviolet effects become less important as the temporal separation increases. 
Overall, the figure suggests that the temperature dependence primarily enters through the overall normalization of the correlator, whereas the 
qualitative $\tau T$-dependence remains similar for both ensembles.\\

 \begin{figure}[h!]
\scalebox{0.70}{\includegraphics{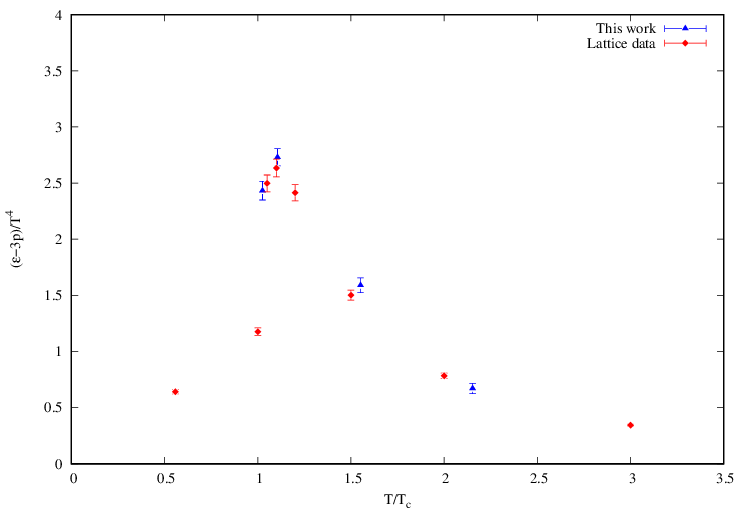}} 
\caption{\label{fig7}  The trace anomaly for improved EMT. The reference lattice data values are taken from continuum-extrapolated SU(3) lattice 
equation-of-state results and their Padé interpolation, with the discontinuity at $T_{c}$ corresponding to the latent heat of the first-order transition \cite{Borsanyi2012,Borsanyi2014}.}
\end{figure}

Figure  \ref{fig7}  shows the comparison of temperature dependence of the trace anomaly of the present EMT results with reference lattice data \cite{Borsanyi2012,Borsanyi2014,Meyer2011} at $N_{\tau}=16$.
The trace anomaly rises rapidly from low temperatures as the system approaches the deconfinement region, reaching a maximum around $T/T_{c}= 1.0 - 1.2$..
This peak reflects the strongest breaking of conformal symmetry and the dominance of nonperturbative dynamics near the transition. The trace anomaly decreases 
steadily at higher temperatures, indicating that the system gradually approaches the weakly coupled, approximately conformal regime where $\epsilon \equiv 3p$.  
The improved EMT results reproduce both the peak location and magnitude and show good overall agreement with the lattice data across the full temperature range, 
with only minor deviations at intermediate temperatures. This behaviour is consistent with the expected thermodynamics of the gluon plasma, where nonconformality 
is maximal near $T_{c}$ and diminishes at higher temperatures \cite{Borsanyi2012,Giusti2017}.

\section{Summary and Conclusion}

We have presented a comprehensive study of the renormalization of the gluonic energy-momentum tensor at the one-loop level, employing an improved clover discretization of the field-strength tensor in pure SU(3) lattice gauge theory. In addition to controlling the lattice artifacts, the tadpole improvement and multi-loop clover discretisation play a crucial role in significantly stabilising the perturbative expansion by absorbing large renormalizations into powers of the mean-field factor $u_{0}$, thus rendering the one-loop diagrams ultraviolet-finite after tadpole subtraction. Notably, in Landau gauge and with matching to the continuum $\overline{\mathrm{MS}}$ scheme via a symmetric subtraction procedure, all one-loop contributions are systematically reduced to a minimal basis of scalar Brillouin-zone integrals, thus giving closed-form expressions for the lattice finite coefficient $\mathfrak{B}_{\mathrm{lat}}$ and the renormalization factor $Z_{T}(u_{0})$. This process clearly separated the traceless spin-2 sector from the scalar trace channel and fixed the normalisation of the spin-2 component of the EMT. The decomposition of the loop contributions of  into continuum-like ``core'' terms and lattice-specific ``transport'' corrections provides  a transparent understanding of how discretization effects enter and are systematically suppressed.

Furthermore, we have shown that renormalization has a nontrivial impact on Euclidean correlators, particularly in the 00-channel.. The correlator in the 00-channel receives contributions from the mixed and scalar channels arising from the trace anomaly. This modifies both the normalization and the short-distance structure of the correlator, with direct implications for the extraction of thermodynamic quantities such as the specific heat. Additionally, we demonstrate that the trace anomaly arises solely from the scalar operator and is unaffected by the spin-2 renormalization. By projecting onto the transverse symmetric-traceless channel, we obtained the renormalization constant.

The consistency of the renormalization procedure and the validity of the operator basis were confirmed by the improved lattice construction of the EMT. Our results show that the improved EMT framework accurately matches the expected thermodynamic behaviour, including the trace anomaly and its temperature dependence, when compared to available lattice results. The deviations from the non-perturbative results can be attributed to higher-order corrections and finite lattice-spacing effects. This suggests that the current one-loop calculation provides a reliable baseline for further quantitative analyses. Anticipated extensions to full QCD with dynamical fermions at two-loop order, and to integration with nonperturbative renormalization schemes such as gradient flow, are likely to further improve the  applicability of this procedure.

\section{Acknowledgements}
ML thankfully acknowledges the computer resources provided by the  National Supercomputing Center, Zhengzhou. 


\end{document}